\documentclass[a4paper]{jpconf}
\usepackage{graphicx}

\def\ltap{\ \raisebox{-.4ex}{\rlap{$\sim$}} \raisebox{.4ex}{$<$}\ }

\begin{document}
\title{Sterile neutrino decay and the LSND experiment}

\author{Sergio Palomares-Ruiz}

\address{Department of Physics and Astronomy, Vanderbilt
University, Nashville, Tennessee 37235-1807}

\ead{sergio.palomares-ruiz@vanderbilt.edu}

\begin{abstract}
We propose a new explanation of the intriguing LSND evidence for
electron antineutrino appearance in terms of heavy (mostly sterile)
neutrino decay via a coupling with a light scalar and light (mostly
active) neutrinos. We perform a fit to the LSND data, as well as all
relevant null-result experiments, taking into account the distortion
of the spectrum due to decay. By requiring a coupling $g \sim
10^{-5}$, a heavy neutrino mass $m_4 \sim 100$~keV and a mixing with
muon neutrinos $|U_{\mu 4}|^2 \sim 10^{-2}$, we show that this model
explains all existing data evading constraints that disfavor standard
(3+1) neutrino models.
\end{abstract}

\vspace{-7mm} 


The confirmation of the solar~\cite{solar} and atmospheric~\cite{atm}
neutrino oscillations by reactor~\cite{KAM} and accelerator~\cite{K2K}
experiments and the bounds from laboratory neutrino mass
measurements~\cite{mass} have set the ``standard'' neutrino picture in
terms of three light active neutrinos with two mass square
differences. However, the observed excess of $\bar{\nu}_e$ in the LSND
experiment~\cite{LSND} was interpreted as evidence of $\bar\nu_\mu
\rightarrow \bar\nu_e$ transitions with a mass square splitting,
$\Delta m^{2}_{LSND} \sim 1 -10 \, \rm{eV}^2$, that cannot be
accommodated by the other two, $\Delta m^{2}_{sol} \sim 8 \times
10^{-5} \, \rm{eV}^2$ and $\Delta m^{2}_{atm} \sim 2.5 \times 10^{-3}
\, \rm{eV}^2$. The most natural solution would be to introduce a
fourth neutrino~\cite{sterile} with the appropriate mass to give the
hinted mass square difference. However, this does not turn out to lead
to a satisfactory description of all data in terms of neutrino
oscillations~\cite{4nusfits} because of tight constraints from
solar~\cite{solar}, atmospheric~\cite{atm}, and null-result
short-baseline (SBL) experiments~\cite{SBL}. In view of this, during
the last years, other alternative explanations have been
proposed~\cite{Ma,otherexpl}. Unlike in Ref~\cite{Ma} where decay was
invoked to evade SBL results, here we propose (see Ref.~\cite{PPS05}
for details) a new explanation of LSND in terms of heavy (mostly
sterile) neutrino decay. This new massive state, $n_4$, must have a 
small mixing with the muon neutrino, $U_{\mu 4}$, and the signal is
explained by its decay into a combination of light neutrinos, being
predominantly of electron type.


A natural way to introduce neutrino decay is by means of a term in
the Lagrangian, which couples neutrinos to a light scalar. We assume
for the scalar mass $m_{1,2,3} \ltap m_\phi \ll m_{4}$, where
$m_{1,2,3}$, $m_\phi$ and $m_4$ are the masses of three light
neutrinos, of the scalar and of the new massive state,
respectively. In this way the three light neutrinos are stable. Hence,
the terms in the Lagrangian which provide the decay are given, in the
mass basis, by
\begin{equation}
\label{eq:Ldec}
\mathcal{L} = - 
\sum_{l} g_{4l} \, \overline{\nu}_{lL} \, n_{4R} \, \phi  +
\mathrm{h.c.}\,, 
\end{equation}
where $l = 1,2,3$ and, in general, the coupling matrix $g_{4l}$ is
complex.

In the case of Majorana particles, neutrinos are identical to
antineutrinos. Weak interactions couple to left-handed (chiral)
neutrinos and right-handed (chiral) antineutrinos while the
propagation states are those of definite helicity. In the relativistic
case, one can identify neutrinos and antineutrinos with helicity
states up to terms of order $m/E_\nu$. It follows from
Eq.~(\ref{eq:Ldec}) that not only the decay $n \to \bar{\nu} + \phi$
can occur, but also $n \to \nu + \phi$ is
possible~\cite{decaywidth}. Hence, the expected number of $\bar\nu_e$
events with neutrino energy in the interval $[E_{\bar\nu_e} ,
  E_{\bar\nu_e}+ dE_{\bar\nu_e}]$ is given by~\cite{PPS05}
\begin{equation}
\label{eq:N}
\frac{dN}{dE_{\bar\nu_e}} = C \, \sigma(E_{\bar\nu_e}) \, 
\left[
\phi_0 
\frac{dP_{\nu_\mu \to \bar\nu_e}(E_{\nu_\mu}^{(\pi)})}{dE_{\bar\nu_e}}
+
\int_{E_{\bar\nu_e}}^{E_{\bar\nu_\mu}^\mathrm{max}}
dE_{\bar\nu_\mu} \, \phi_\mu(E_{\bar\nu_\mu}) \,
\frac{dP_{\bar\nu_\mu \to \bar\nu_e}(E_{\bar\nu_\mu})}{dE_{\bar\nu_e}}
\right]\,,
\end{equation}
where $\phi_\mu(E_{\bar\nu_\mu})$ is the muon $\bar{\nu}_\mu$
spectrum, $\phi_0 = \int dE_{\bar\nu_\mu} \phi_\mu(E_{\bar\nu_\mu})$
and $dP_{\nu_\alpha^r\to\nu_\beta^s}(E_{\nu_\alpha})/dE_{\nu_\beta}$
is the differential probability that a neutrino of flavor $\alpha$
with energy $E_{\nu_\alpha}$ is converted into an neutrino of flavor
$\beta$ with energy in the interval $[E_{\nu_\beta}, E_{\nu_\beta} +
  dE_{\nu_\beta}]$. The indexes $r,s$ take the value `$-$' (`$+$') for
neutrinos (antineutrinos). The detection cross section is given by
$\sigma(E_{\bar\nu_e})$ and $C$ is an overall constant containing the
number of target particles, efficiencies and geometrical factors.

Unlike other explanations of the LSND result based on sterile
neutrinos, our model does not require mixing of the electron neutrino
with the heavy mass state, so for simplicity we assume $U_{e4} =
0$. Hence the differential probability for $\nu_\mu^r \rightarrow
\nu_e^s$ conversion can be written as~\cite{PPS05}
\begin{equation}
\label{eq:P2}
\frac{dP_{\nu_\mu^r\to \nu_e^s}(E_{\nu_\mu})}{dE_{\nu_e}} =
 |U_{\mu 4}|^2 \, (1 - e^{-\Gamma_4 L}) \,
\Theta(E_{\nu_\mu} - E_{\nu_e}) \times
\left\{
\begin{array}{l@{\quad}l}
E_{\nu_e}/E_{\nu_\mu}^2 & r=s , \\
(E_{\nu_\mu} - E_{\nu_e})/E_{\nu_\mu}^2 & r \neq s ,
\end{array} \right.,
\end{equation}
where $\Gamma_4$ is the decay rate of $n_4$ as follows from
Eq.~(\ref{eq:Ldec})~\cite{decaywidth}.


\begin{figure}[t]
\begin{minipage}{18pc}
\begin{center}
\includegraphics[width=5.5cm,height=4.5cm]{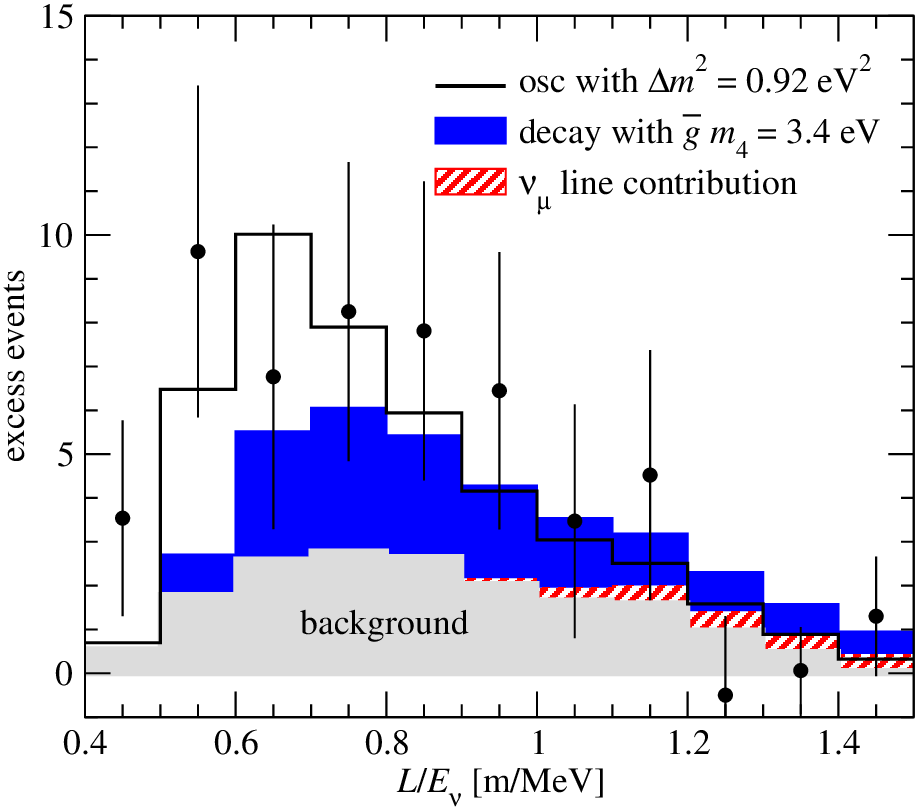}
\end{center}
\vspace{-2mm}
\caption{\label{fig:spectrum} \footnotesize $L/E_\nu$ spectrum for
  LSND for decay and oscillations compared with the data given in
  Fig.~24 of Ref.~\cite{LSND}. The hatched histogram shows the
  contribution of the $\nu_\mu$ line from the primary pion decay in
  the decay scenario. From Ref.~\cite{PPS05}.} 
\end{minipage}\hspace{2pc}%
\begin{minipage}{18pc}
\begin{center}
\vspace{-2mm}
\includegraphics[width=6.45cm,height=5.03cm]{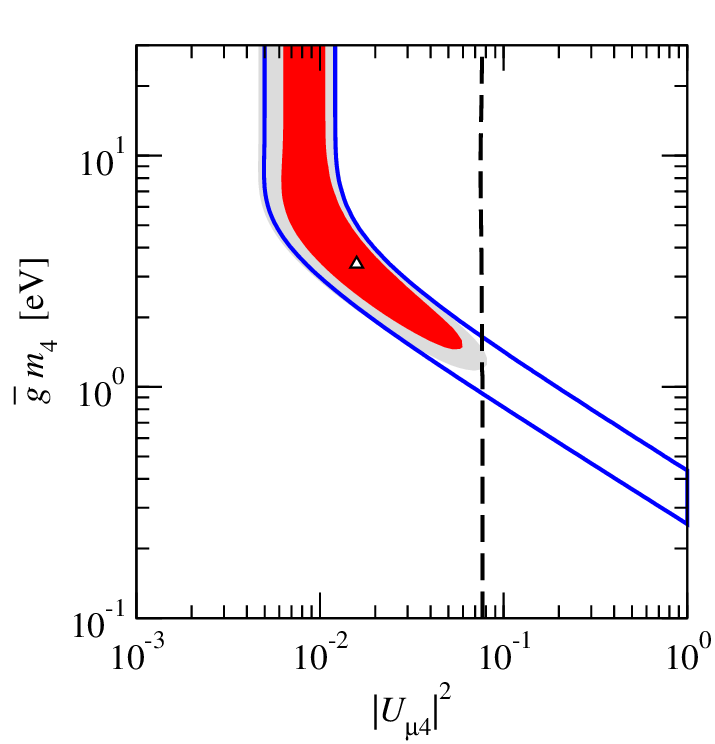}
\end{center}
\vspace{-6mm}
\caption{\label{fig:analysis} \footnotesize Allowed regions for LSND +
  KARMEN (solid) and SBL disappearance+atmospheric neutrino experiments
  (dashed) at 99\%~CL, and their combination (shaded regions) at 90\%
  and 99\%~CL. From Ref.~\cite{PPS05}.}
\end{minipage} 
\vspace{-6mm}
\end{figure}

Our analysis of LSND data (Fig.~\ref{fig:spectrum}), which does not
use the much less significant decay in flight (DIF) sample, gives a
best fit value of $\chi^2_\mathrm{min} = 5.6/9$~dof for oscillations
and $\chi^2_\mathrm{min} = 10.8/9$~dof for decay. The reason for the
slightly worse fit for decay is the spectral distortion implied by the
energy distribution of the decay products. Nevertheless, the overall
goodness of fit (GOF) is still acceptable, 29\%, giving as best fit
values $|U_{\mu 4}|^2 = 0.016$ and $\bar g \, m_4 = 3.4\,\mathrm{eV}$,
where $\bar g^2 \equiv \sum_l |g_{4l}|^2$. 

As is well known, (3+1) neutrino models are disfavored due to the
tension between LSND and SBL experiments, which have not observed
either appearance or dissapearance so far~\cite{SBL}. We have also
studied their compatibility in the decay scenario. Under the
assumption $U_{e4}=0$, no $\bar\nu_e$ disappearance is expected in SBL
reactor experiments. However, since the explanation of the LSND signal
within this scenario requires mixing of $\nu_\mu$ with the heavy
state, one expects some effect in $\nu_\mu$ disappearance
experiments. The main constraints on the allowed LSND-KARMEN region
come from the CDHS experiment~\cite{SBL} and from atmospheric neutrino
data~\cite{atm4nu}. Nevertheless, the different expression for the
$\nu_\mu$ survival probability in the decay scenario provides a rather
poor sensitivity to $U_{\mu 4}$, which comes via $|U_{\mu 4}|^4$,
unlike the case of oscillations, that it comes via $|U_{\mu
  4}|^2$. The results are shown in Fig.~\ref{fig:analysis}, where
appearance and disappearance experiments are shown to be in perfect
agreement, with large overlap of the allowed regions.


Mixing between active and sterile neutrinos, and couplings between
active neutrinos and a light scalar, have been extensively studied,
both in laboratory experiments and, for their implications in the
evolution of the early Universe and of supernovae. The decay of pions
and kaons has been used to set bounds on the mixing of heavy
neutrinos for masses above MeV~\cite{KPS04}. In our model, the
most stringent bounds on the coupling from laboratory experiments,
coming from light scalar emission in pion and kaon decays, are of
order $\bar{g}^2 < \mathrm{few} \times 10^{-5}$~\cite{coupling}. On
the other hand, a coupling $\bar{g} \sim 10^{-5}$ would imply that
$n_4$ and $\phi$ are strongly coupled to active neutrinos inside the
supernova, and hence, they are trapped within the neutrinosphere,
avoiding any energy loss due to particles escaping from the core. This
value of the coupling would also imply that as soon as $n_4$ are
produced in the Early Universe, they are thermalized along with the
scalar, leading to 1.57 extra number of relativistic degrees of
freedom at BBN.

In summary, we have presented an explanation of the LSND evidence for
$\bar\nu_e$ appearance based on the decay of a heavy (mostly sterile)
neutrino into a light scalar particle and light neutrinos. It is
needed $|U_{\mu 4}|^2 \sim 10^{-2}$ and $\bar{g} m_4 \sim 1$ eV, and 
hence for $\bar{g} \sim 10^{-5}$, $m_4 \sim 100$ keV. In addition, the
decay model also predicts a detectable signal in the upcoming
MiniBooNE experiment~\cite{miniboone}.

\vspace{-2mm}

\ack 

These results were obtained in collaboration with S. Pascoli and
T. Schwetz. SPR is supported by NASA Grant ATP02-0000-0151 and by the
Spanish Grant FPA2002-00612 of the MCT.


\section*{References}

\begin{thebibliography}{99}


\bibitem{solar}
  Q.~R.~Ahmad {\it et al.}  
  2002 {\it Phys.\ Rev.\ Lett.\ } {\bf 89} 011301


\bibitem{atm} 
  Ashie Y {\it et al.} 
  2005 {\it Phys.\ Rev.\ } D {\bf 71} 112005


\bibitem{KAM}
  Araki T {\it et al.}  
  2005 {\it Phys.\ Rev.\ Lett.\ } {\bf 94} 081801


\bibitem{K2K}
  Aliu E {\it et al.} 
  2005 {\it Phys.\ Rev.\ Lett.\ } {\bf 94} 081802


\bibitem{mass}
  Weinheimer C
  2003 {\it Nucl.\ Phys.\ Proc.\ Suppl.\ } 118 279


\bibitem{LSND} 
  Aguilar A {\it et al.}  
  2001 {\it Phys.\ Rev.\ } D {\bf 64} 112007


\bibitem{sterile}
  Peltoniemi J T, Tommasini D and Valle J W F
  1993 {\it Phys.\ Lett.\ } B {\bf 298} 383

\nonum
  Caldwell D O and Mohapatra R N
  1993 {\it Phys.\ Rev.\ } D {\bf 48} 3259


\bibitem{4nusfits}
  Strumia A
  2002 {\it Phys.\ Lett.\ B} {\bf 539} 91

\nonum
  Maltoni M, Schwetz T, Tortola M A and Valle J W F
  2004 {\it New J.\ Phys.\ } {\bf 6} 122


\bibitem{SBL}
  Armbruster B {\it et al.}  
  2002 {\it Phys.\ Rev.\ } D {\bf 65} 112001

\nonum
  Dydak F {\it et al.}
  1984 {\it Phys.\ Lett.\ } B {\bf 134} 281

\nonum
  Declais Y {\it et al.}
  1995 {\it Nucl.\ Phys.\ } B {\bf 434} 503

\nonum
  Astier P {\it et al.}  
  2003 {\it Phys.\ Lett.\ } B {\bf 570} 19

\bibitem{Ma}
  Ma E, Rajasekaran G and Stancu I
  2000 {\it Phys.\ Rev.\ } D {\bf 61} 071302 


\bibitem{otherexpl}
  Murayama H and Yanagida T
  2001 {\it Phys.\ Lett.\ } B {\bf 520} 263

\nonum
  Barenboim G, Borissov L and Lykken J 
  2002 {\it Preprint} hep-ph/0212116

\nonum
  Peres O L G and Smirnov A Y
  2001 {\it Nucl.\ Phys.\ } B {\bf 599} 3

\nonum
  Sorel M, Conrad J M and Shaevitz M
  2004 {\it Phys.\ Rev.\ } D {\bf 70} 073004

\nonum
  Babu K S and Pakvasa S
  2002 {\it Preprint} hep-ph/0204236

\nonum
  Barger V, Marfatia D and Whisnant K
  2003 {\it Phys.\ Lett.\ } B {\bf 576} 303

\nonum
  Kaplan D B, Nelson A E and Weiner N
  2004 {\it Phys.\ Rev.\ Lett.\ } {\bf 93} 091801 

\nonum
  Gelmini G, Palomares-Ruiz S and Pascoli S
  2004 {\it Phys.\ Rev.\ Lett.\  } {\bf 93} 081302

\nonum
  Barenboim G and Mavromatos N E
  2005 {\it JHEP } {\bf 0501} 034

\nonum
  Kostelecky V A and Mewes M
  2004 {\it Phys.\ Rev.\ } D {\bf 70} 076002

\nonum
  Paes H, Pakvasa S and Weiler T J
  2005 {\it Phys.\ Rev.\ } D {\bf 72} 095017


\bibitem{PPS05}
  Palomares-Ruiz S, Pascoli S and Schwetz T
  2005 {\it JHEP} {\bf 0509} 048


\bibitem{decaywidth}
  Kim C W and Lam W P
  1990 {\it Mod.\ Phys.\ Lett.\ } A {\bf 5} 297

\nonum
  Giunti C, Kim C W, Lee U W and Lam W P
  1992 {\it Phys.\ Rev.\ } D {\bf 45} 1557


\bibitem{atm4nu}
  Maltoni M, Schwetz T and Valle J W F
  2002 {\it Phys.\ Rev.\ } D {\bf 65} 093004 


\bibitem{KPS04}
  Kusenko A, Pascoli S and Semikoz D
  2005 {\it JHEP} {\bf 0511} 028


\bibitem{coupling}
  Britton D I {\it et al.}
  1994 {\it Phys.\ Rev.\ } D {\bf 49} 28

\nonum
  Barger V D, Keung W Y and Pakvasa S
  1982 {\it Phys.\ Rev.\ } D {\bf 25} 907

\nonum
  Gelmini G B, Nussinov S and Roncadelli M
  1982 {\it Nucl.\ Phys.\ } B {\bf 209} 157


\bibitem{miniboone} 
  The MiniBooNE Run Plan, available at
  \small{\texttt{http://www-boone.fnal.gov/publicpages/}}


\end{thebibliography}
\end{document}